\documentclass[doublespacing,review]{elsarticle}
\usepackage{amssymb}
\usepackage{amsmath}
\usepackage{amsthm}
\usepackage{graphics}

\journal{Annals of Physics}

\begin{document}

\begin{frontmatter}



\title{Full counting statistics as a probe of quantum coherence in a
side-coupled double quantum dot system}

\author{Hai-Bin Xue\corref{cor1}}
\cortext[cor1]{Corresponding author. Tel.: +86 351 6018030; fax: +86
351 6018030.} \ead {xuehaibin@tyut.edu.cn}
\address{College of Physics and Optoelectronics, Taiyuan University of Technology,
Taiyuan 030024, China}

\begin{abstract}

We study theoretically the full counting statistics of electron
transport through side-coupled double quantum dot (QD) based on an
efficient particle-number-resolved master equation. It is
demonstrated that the high-order cumulants of transport current are
more sensitive to the quantum coherence than the average current,
which can be used to probe the quantum coherence of the considered
double QD system. Especially, the quantum coherence plays a crucial
role in determining whether the super-Poissonian noise occurs in the
weak inter-dot hopping coupling regime depending on the
corresponding dot-lead coupling, and the corresponding values of
super-Poissonian noise can be relatively enhanced when considering
the spins of conduction electrons. Moreover, this super-Poissonian
noise bias range depends on the singly-occupied eigenstates of the
system, which thus suggests a tunable super-Poissonian noise device.
The occurrence-mechanism of super-Poissonian noise can be understood
in terms of the interplay of quantum coherence and effective
competition between fast-and-slow transport channels.

\end{abstract}


\end{frontmatter}

\newpage

\section{Introduction}

Non-equilibrium electronic full counting statistics (FCS) is a
powerful diagnostic tool for probing the nature of electron
transport mechanisms inaccessible by the average current
measurements\cite{Blanter,Nazarov}. Recently, probing the quantum
coherence of coupled quantum-dot (QD) systems by means of FCS, i.e.,
the transport current high-order cumulants, have attracted
considerable attention due to the quantum coherence plays a crucial
role for their application in the field of solid-state quantum
computing, and some interesting current noise characteristics have
been observed or predicted\cite%
{Urban,WangSK,KieBlich01,Groth,KieBlich02,Welack,Fang01,Fang02}. For
example, the interplay of quantum coherence and strong Coulomb
blockade/charging energy can induce the super-Poisson behavior of
transport current in the parallel\cite{Urban,WangSK} and
series\cite{KieBlich01} double-QD (DQD) systems, but for coupled
three-QD system\cite{Groth}, a crossover from sub-Poissonian to
super-Poissonian statistics was observed with increasing ratio of
tunnel and decoherence rates. In particular, the high-order
cumulants, e.g., the shot noise, the skewness, are more sensitive to
the quantum coherence effect than the average current in the
different types of QD systems\cite{KieBlich02,Welack,Fang01,Fang02},
namely, the series DQD\cite{KieBlich02}, the parallel
DQD\cite{Welack}, the Aharonov-Bohm interferometer with a quantum
dot embedded in one of its two
current paths\cite{Fang01}, a semiopen Kondo-correlated quantum dot\cite%
{Fang02}. Moreover, theoretical studies have also shown that the
high-order cumulants can be used to detect the positions of the
zeros of the generating function\cite{Kambly}, reveal the intrinsic
multistability\cite{Schaller}, and extract the fractional charge of
charge transfer through an impurity in a chiral Luttinger
liquid\cite{Komnik}. However, the shot noise or higher-order
cumulants of the current do not provide any more information on
decoherence additional than the average current that also was
observed in two types of mesoscopic structures with a varying number
of propagating channels, e.g.,\ mesoscopic cavities and
Aharonov-Bohm rings\cite{Pala}. Therefore, extracting quantum
coherence information from the current high-order cumulants is still
an open issue. On the other hand, the FCS of electron transport
through a semiconductor quantum dot (QD) have been experimentally,
especially the fifteen-order cumulants\cite{Flindt,Fricke} and
finite-frequency current statistics\cite{Ubbelohde} can be extracted
from the high-quality real-time single-electron measurements. This
provides the opportunity to investigate the relationship between the
quantum coherence and the FCS of electron transport through coupled
QD systems.

The goal of this paper is thus to study the influence of the quantum
coherence on the high-order cumulants of electron transport through
a relatively highly coherent quantum system, and analyze the
feasibility of extracting quantum coherence information from the
current high-order cumulants. Here, we consider a side-coupled DQD
system with high quantum coherent for a weak inter-dot hopping
coupling relative to the dot-lead
coupling. Although super-Poissonian shot noise was studied in this QD system%
\cite{Weymann,Djurica}, the effects of the quantum coherence between
singly-occupied eigenstates of the system, i.e. the off-diagonal
elements of the reduced density matrix, and the electron spin on
super-Poissonian noise, especially the feasibility of probing the
quantum coherence of this QD
system by means of FCS, are not revealed in previous investigations\cite%
{Djurica}. Here, we examine the effects of conduction electron spin
and quantum coherence on the FCS in such QDs system. We found that
the current high-order cumulants are more sensitive to the quantum
coherence than the average current, which can be used to probe the
quantum coherence of this side-coupled DQD system. Especially, the
quantum coherence of this DQD system plays a crucial role in
determining whether the super-Poissonian noise occurs in the weak
inter-dot hopping coupling regime, and the corresponding values of
super-Poissonian noise can be relatively enhanced when considering
conduction electron spin. Moreover, this super-Poissonian noise
characteristic, which originates from the quantum coherence between
the singly-occupied eigenstates, can be used to design a tunable
super-Poissonian noise device because of singly-occupied eigenstates
can be tuned by a gate voltage. The paper is organized as follows.
In Sec. II, we introduce the considered side-coupled DQD system and
outline the procedure to obtain the FCS formalism based on an
effective particle-number-resolved quantum master equation approach.
The numerical results are discussed in Sec. III, where we discuss
the effects of the conduction electron spin and quantum coherence on
super-Poissonian noise and analyze the mechanism of their formation.
Finally, in Sec. IV we summarize the work.

\section{MODEL AND FORMALISM}

We consider a side-coupled DQD system weakly coupled to two metallic
electrodes [see the Fig. 1]. The system is described by the Hamiltonian $%
H_{total}=H_{dot}+H_{leads}+H_{T}$. The double-QD Hamiltonian is
given by
\begin{equation}
H_{dot}=\sum_{i\sigma }\epsilon _{i}d_{i\sigma }^{\dag }d_{i\sigma
}+U_{12}\sum_{\sigma \sigma ^{\prime }}\hat{n}_{1\sigma
}\hat{n}_{2\sigma ^{\prime }}-J\,\sum_{\sigma }\left( d_{1\sigma
}^{\dag }d_{2\sigma }+d_{2\sigma }^{\dag }d_{1\sigma }\right) ,
\label{model}
\end{equation}%
where $d_{i\sigma }^{\dag }$ ($d_{i\sigma }$) creates (annihilates)
an electron with spin $\sigma $ and energy $\varepsilon _{i}$ (which
can be tuned by a gate voltage $V_{g}$) in $i$th QD. $U_{12}$ is the
interdot Coulomb repulsion between two electrons in the DQD system,
where we consider the intradot Coulomb interaction $U\rightarrow
\infty ,$ so that the double-electron occupation in the same QD is
prohibited but in different QDs is permitted. The last term of
$H_{dot}$ describes the hopping coupling between the two dots with
$J\,\ $being the hopping parameter.

The relaxation in the electrodes is assumed to be sufficiently fast
so that their electron distributions can be described by equilibrium
Fermi functions. The electrodes are modeled as non-interacting Fermi
gases and the corresponding Hamiltonian%
\begin{equation}
H_{Leads}=\sum_{\alpha \mathbf{k}\sigma }\varepsilon _{\alpha \mathbf{k}%
}a_{\alpha \mathbf{k}\sigma }^{\dag }a_{\alpha \mathbf{k}\sigma },
\label{Leads}
\end{equation}%
\ \ where $a_{\alpha \mathbf{k}}^{\dag }$ ($a_{\alpha \mathbf{k}}$)
creates (annihilates) an electron with energy $\varepsilon _{\alpha
\mathbf{k}}$ and momentum $\mathbf{k}$ in $\alpha $ ($\alpha =L,R$)
electrode. The tunneling between the QD-1 and the electrodes is
described by
\begin{equation}
H_{T}=\sum_{\alpha \mathbf{k}\sigma }\left( t_{\alpha
\mathbf{k}}a_{\alpha \mathbf{k}\sigma }^{\dag }d_{1\sigma
}+t_{\alpha \mathbf{k}}^{\ast }d_{1\sigma }^{\dag }a_{\alpha
\mathbf{k}\sigma }\right) .  \label{tunneling}
\end{equation}

The QD-electrode coupling is assumed to be sufficiently weak, such
that the sequential tunneling is dominant. The transitions are well
described by quantum master equation of a reduced density matrix
spanned by the eigenstates of the side-coupled DQD system. The
detailed derivation of the FCS formalism based on the
particle-number-resolved quantum master equation can be found in
Refs. \cite{WangSK,Li01,Li02}, and here we only give the main
results. Under the second order Born approximation and Markov
approximation, the particle-number-resolved quantum master equation
for the reduced density matrix is given by
\begin{equation}
\dot{\rho}^{\left( n\right) }\left( t\right) =-i\mathcal{L}\rho
^{\left( n\right) }\left( t\right) -\frac{1}{2}\mathcal{R}\rho
^{\left( n\right) }\left( t\right) ,  \label{Master1}
\end{equation}%
with%
\begin{eqnarray*}
\mathcal{R}\rho ^{\left( n\right) }\left( t\right) & =\sum_{\sigma
=\uparrow ,\downarrow }\left[ d_{1\sigma }^{\dagger }A_{L,1\sigma
}^{\left( -\right) }\rho ^{\left( n\right) }\left( t\right)
+d_{1\sigma }^{\dagger }A_{R,1\sigma }^{\left( -\right) }\rho
^{\left( n\right) }\left( t\right)\right.   \\
& +\rho ^{\left( n\right) }\left( t\right) A_{L,1\sigma }^{\left(
+\right) }d_{1\sigma }^{\dagger }+\rho ^{\left( n\right) }\left(
t\right)A_{R,1\sigma }^{\left( +\right) }d_{1\sigma }^{\dagger }  \\
& -A_{L,1\sigma }^{\left( -\right) }\rho ^{\left( n\right) }\left(
t\right) d_{1\sigma }^{\dagger }-A_{R,1\sigma }^{\left( -\right)
}\rho ^{\left(n-1\right) }\left( t\right) d_{1\sigma }^{\dagger }  \\
& \left. -d_{1\sigma }^{\dagger }\rho ^{\left( n\right) }\left(
t\right) A_{L,1\sigma }^{\left( +\right) }-d_{1\sigma }^{\dagger
}\rho ^{\left( n+1\right) }\left( t\right) A_{R,1\sigma }^{\left(
+\right) }\right] +H.c., \label{Master2}
\end{eqnarray*}%
where $A_{\alpha ,1\sigma }^{\left( \pm \right) }=\Gamma _{\alpha
}n_{\alpha }^{\left( \pm \right) }\left( -\mathcal{L}\right)
d_{1\sigma },n_{\alpha }^{+}=f_{\alpha },n_{\alpha }^{-}=1-f_{\alpha
}$ ($f_{\alpha }$ is the Fermi function of the electrode $\alpha $),
and $\Gamma _{\alpha =L,R}=2\pi g_{\alpha =L,R}\left\vert t_{\alpha
=L,R}\right\vert ^{2}$. Liouvillian
superoperator $\mathcal{L}$ is defined as $\mathcal{L}\left( \cdots \right) =%
\left[ H_{dot},\left( \cdots \right) \right] $, and $g_{\alpha
=L,R}$ denotes the density of states of the metallic electrodes.
$\rho ^{\left( n\right) }\left( t\right) $ is the reduced density
matrix of the side-coupled DQD system conditioned by the electron
numbers arriving at the right electrode up to time $t$. In order to
calculate the FCS, one can define $S\left( \chi ,t\right)
=\sum_{n}\rho ^{\left( n\right) }\left( t\right) e^{in\chi }$.
According to the definition of the\textbf{\ }cumulant
generating function\cite{Bagrets} $e^{-F\left( \chi \right) }=\sum_{n}$Tr$%
\left[ \rho ^{\left( n\right) }\left( t\right) \right] e^{in\chi
}=\sum_{n}P\left( n,t\right) e^{in\chi }$, we evidently have
$e^{-F\left( \chi \right) }=$Tr$\left[ S\left( \chi ,t\right)
\right] $, where the trace is over the eigenstates of the considered
DQD system and $\chi $ is the counting field. Since Eq.
(\ref{Master1}) has the following form
\begin{equation}
\dot{\rho}^{\left( n\right) }=A\rho ^{\left( n\right) }+C\rho
^{\left( n+1\right) }+D\rho ^{\left( n-1\right) },
\label{formalmaster}
\end{equation}%
$S\left( \chi ,t\right) $ satisfies
\begin{equation}
\dot{S}=AS+e^{-i\chi }CS+e^{i\chi }DS\equiv \mathcal{L}_{\chi }S,
\label{formalmaster1}
\end{equation}%
where $S$ is a column matrix, and $A$, $C$ and $D$ are three square
matrices. In the low frequency limit, the counting time ($i.e.$, the
time of measurement) is much longer than the time of electron
tunneling through the
considered DQD system. In this case, $F\left( \chi \right) $ is given by\cite%
{Groth,Bagrets}
\begin{equation}
F\left( \chi \right) =-\lambda _{1}\left( \chi \right) t,
\label{CGFformal}
\end{equation}%
where $\lambda _{1}\left( \chi \right) $ is the eigenvalue of $\mathcal{L}%
_{\chi }$ which goes to zero for $\chi \rightarrow 0$. According to
the definition of the cumulants one can express $\lambda _{1}\left(
\chi \right) $\ as
\begin{equation}
\lambda _{1}\left( \chi \right) =\frac{1}{t}\sum_{k=1}^{\infty }C_{k}\frac{%
\left( i\chi \right) ^{k}}{k!}.  \label{Lambda}
\end{equation}%
Here, the first four cumulants $C_{k}$ are directly related to the
transport characteristics. For example, the first-order cumulant
(the peak position of the distribution of transferred-electron
number) $C_{1}=\bar{n}$ gives the average current $\left\langle
I\right\rangle =eC_{1}/t$. The zero-frequency shot noise is related
to the second-order cumulant (the peak-width of the
distribution) $S=2e^{2}C_{2}/t=2e^{2}\left( \overline{n^{2}}-\bar{n}%
^{2}\right) /t$. The third-order $C_{3}=\overline{\left(
n-\bar{n}\right) ^{3}}$ and fourth-order cumulants
$C_{4}=\overline{\left( n-\bar{n}\right) ^{4}}$, respectively,
characterize the skewness and kurtosis of the distribution. Here,
$\overline{\left( \cdots \right) }=\sum_{n}\left( \cdots \right)
P\left( n,t\right) $. In general, the shot noise, skewness and
kurtosis are represented by the Fano factor $F_{2}=C_{2}/C_{1}$, $%
F_{3}=C_{3}/C_{1}$ and $F_{4}=C_{4}/C_{1}$, respectively. Moreover,
the specific form of $\mathcal{L}_{\chi }$ can be obtained by
performing a
discrete Fourier transformation to the matrix elements of Eq. (\ref{Master1}%
). Inserting Eq. (\ref{Lambda}) into $\left\vert \mathcal{L}_{\chi
}-\lambda _{1}\left( \chi \right) I\right\vert =0$ and expanding
this determinant in series of $\left( i\chi \right) ^{k}$, one can
calculate $C_{k}/t$ by setting the coefficient of $(i\chi )^{k}$
equal to zero.

\section{NUMERICAL RESULTS AND DISCUSSION}

We now study the effects of conduction-electron spin and quantum
coherence on the FCS of electronic transport through the
side-coupled DQD system
weakly coupled to two metallic electrodes. We assume the bias voltage ($%
V_{b}=\mu _{L}-\mu _{R}$) is symmetrically entirely dropped at the
QD-electrode tunnel junctions, which implies that the levels of the
QDs are independent of the applied bias voltage even if the
couplings are not symmetric, and choose meV as the unit of energy
which corresponds to a
typical experimental situation\cite{Elzerman}. Here, the eigenstates of $%
H_{dot}$ are chosen to describe the electronic states of this
side-coupled double-QD system, which can be obtained by
diagonalizing the QDs Hamiltonian $H_{dot}$ in the basis represented
by the electron spins in the QD-1 and QD-2 denoted respectively by
$\left\vert \sigma _{1},\sigma _{2}\right\rangle $, i.e., $\left\{
\left\vert 0,0\right\rangle ,\left\vert \uparrow ,0\right\rangle
,\left\vert \downarrow ,0\right\rangle ,\left\vert 0,\uparrow
\right\rangle ,\left\vert 0,\downarrow \right\rangle ,\left\vert
\uparrow ,\uparrow \right\rangle ,\left\vert \downarrow ,\uparrow
\right\rangle ,\left\vert \uparrow ,\downarrow \right\rangle
,\left\vert \downarrow ,\downarrow \right\rangle \right\} $. In the
present work, we
only study the transport above the sequential tunneling threshold, i.e., $%
V_{b}>2\epsilon _{se}$, where $\epsilon _{se}$\ is the energy
difference
between the ground state with charge $N$\ and the first excited states $N-1$%
\cite{Aghassi}. In this regime, the inelastic sequential tunneling
process is dominant. It should be noted that, however, the
normalized second-, third- and fourth-order cumulants will deviate
from the results obtained by
considering only sequential tunneling when taking into account cotunneling%
\cite{Thielmann}, since in the Coulomb blockade regime the current
is exponentially suppressed and the electron transport is dominated
by cotunneling. In the following numerical calculations, the
parameters of the
QDs are chosen as $\epsilon _{1}=\epsilon _{2}=\epsilon $, $U_{12}=5$, $%
J=0.001$ and $k_{B}T=0.1$.

We firstly consider the influence of the conduction-electron spin on
the FCS of electronic transport. The spinless Hamiltonian of the
considered
double-QD system is given by\cite{Djurica}%
\begin{equation}
H_{spinless}=\sum_{i}\epsilon _{i}d_{i}^{\dag }d_{i}+U_{12}\hat{n}_{1}\hat{n}%
_{2}-J\,\left( d_{1}^{\dag }d_{2}+d_{2}^{\dag }d_{1}\right) .
\label{Spinless}
\end{equation}%
Figure 2 shows the average current, shot noise, skewness and
kurtosis as a function of the bias voltage for the two cases of
considering and without considering the conduction-electron spin.
Here, the off-diagonal elements of the reduced density matrix, i.e.,
quantum coherence, are considered in the numerical calculations. For
the symmetric coupling of the QD-1 with two metallic electrodes,
i.e., $\Gamma _{L\uparrow }=\Gamma _{L\downarrow }=\Gamma
_{R\uparrow }=\Gamma _{R\downarrow }=\Gamma $, the magnitude of
average current is relatively large for considering the
conduction-electron spin case, see the Figs. 2(a) and 2(e). This
feature can be explained in terms of the variation of the
probability of the corresponding eigenstates of the considered DQD
system. For considering and without considering the
conduction-electron spin cases, the transport channel currents can
be
expressed by\cite{Xue01,Xue02,Xue03,Xue04,Timm}%
\begin{equation}
\left\{
\begin{array}{c}
I_{\left\vert \Psi _{1,\sigma }^{\pm }\right\rangle \rightarrow
\left\vert \Psi _{0,0}\right\rangle }=\frac{1}{2}\Gamma _{R\sigma
}n_{R}^{-}\left( \epsilon _{1,\sigma }^{\pm }-\epsilon _{0,0}\right)
P_{\left\vert \Psi
_{1,\sigma }^{\pm }\right\rangle } \\
I_{\left\vert \Psi _{\uparrow ,\uparrow }\right\rangle \rightarrow
\left\vert \Psi _{1,\uparrow }^{\pm }\right\rangle
}=\frac{1}{2}\Gamma _{R\uparrow }n_{R}^{-}\left( \epsilon _{\uparrow
,\uparrow }-\epsilon _{1,\uparrow }^{\pm }\right) P_{\left\vert \Psi
_{\uparrow ,\uparrow
}\right\rangle } \\
I_{\left\vert \Psi _{\downarrow ,\uparrow }\right\rangle \rightarrow
\left\vert \Psi _{1,\uparrow }^{\pm }\right\rangle
}=\frac{1}{2}\Gamma _{R\downarrow }n_{R}^{-}\left( \epsilon
_{\downarrow ,\uparrow }-\epsilon _{1,\uparrow }^{\pm }\right)
P_{\left\vert \Psi _{\downarrow ,\uparrow
}\right\rangle } \\
I_{\left\vert \Psi _{\uparrow ,\downarrow }\right\rangle \rightarrow
\left\vert \Psi _{1,\downarrow }^{\pm }\right\rangle
}=\frac{1}{2}\Gamma _{R\uparrow }n_{R}^{-}\left( \epsilon _{\uparrow
,\downarrow }-\epsilon _{1,\downarrow }^{\pm }\right) P_{\left\vert
\Psi _{\uparrow ,\downarrow
}\right\rangle } \\
I_{\left\vert \Psi _{\downarrow ,\downarrow }\right\rangle
\rightarrow \left\vert \Psi _{1,\downarrow }^{\pm }\right\rangle
}=\frac{1}{2}\Gamma _{R\downarrow }n_{R}^{-}\left( \epsilon
_{\downarrow ,\downarrow }-\epsilon _{1,\downarrow }^{\pm }\right)
P_{\left\vert \Psi _{\downarrow ,\downarrow
}\right\rangle }%
\end{array}%
\right. ,  \label{spincurr}
\end{equation}%
and%
\begin{equation}
\left\{
\begin{array}{c}
I_{\left\vert \Psi _{1}^{\pm }\right\rangle \rightarrow \left\vert
\Psi _{0}\right\rangle }=\frac{1}{2}\Gamma n_{R}^{-}\left( \epsilon
_{1}^{\pm
}-\epsilon _{0}\right) P_{\left\vert \Psi _{1}^{\pm }\right\rangle } \\
I_{\left\vert \Psi _{2}\right\rangle \rightarrow \left\vert \Psi
_{1}^{\pm }\right\rangle }=\frac{1}{2}\Gamma n_{R}^{-}\left(
\epsilon _{2}-\epsilon
_{1}^{\pm }\right) P_{\left\vert \Psi _{2}\right\rangle }%
\end{array}%
\right. ,  \label{spinlesscurr}
\end{equation}%
respectively. Here, $\epsilon _{i}$ is the eigenvalue of the system
eigenstate $\left\vert i\right\rangle $, and $P_{\left\vert
i\right\rangle }$ is the probability of electron occupying state
$\left\vert i\right\rangle $. For the considered system parameters
($\epsilon _{1}=\epsilon _{2}=1$), the first current-step induced by
the transition processes between the singly-occupied and empty
eigenstates, and the Fermi function of the right electrode
$n_{R}^{+}\left( \epsilon _{1,\sigma }^{\pm }-\epsilon _{0,0}\right)
=0$, namely, electrons reverse tunneling from the right
electrode to the considered DQD system is prohibited. This leads to $%
n_{R}^{-}\left( \epsilon _{1,\sigma }^{\pm }-\epsilon _{0,0}\right)
=0$, so that the probability of each eigenstate is given by $1/5$
since there are four singly-occupied and one empty eigenstates in
this bias voltage window, one can check from Eq. (\ref{spincurr})
that the total current $I_{1}/\Gamma =2/5$. The second current-step
induced by the two kinds of transition processes: (i) between the
singly-occupied and empty eigenstates, (ii) between doubly-occupied
and the singly-occupied eigenstates, and the
corresponding Fermi function of the right electrode $n_{R}^{+}=0$, i.e., $%
n_{R}^{-}=0$. In this case, the probability of each eigenstate is
$1/9$ due to there are nine eigenstates in this bias voltage window,
and the total current $I_{2}/\Gamma =2/3$, see the Fig. 2(e). As for
without considering the conduction-electron spin case, based on the
same analysis the magnitudes
of the first current-step and the second current-step are given by $%
I_{1,spinless}/\Gamma =1/3$ and $I_{2,spinless}/\Gamma =1/2$,
respectively, see the Fig. 2(a). Therefore, the current for
considering the conduction-electron spin case is relatively large.
In particular, the corresponding high-order cumulants of transport
current have a significantly strengthened (the shot noise and
skewness) or weakened (the kurtosis) especially for the $\Gamma \gg
J$ case, see the dotted and short-dashed lines in Figs. 2(b) and
2(f), 2(c) and 2(g), 2(d) and 2(h). These characteristics have
demonstrated that the conduction-electron spin cannot be neglect in
the systems with a relatively strong quantum coherent effect.

Now, we discuss the effect of quantum coherence on the high-order
cumulants of transport current, and choose $\epsilon _{1}=\epsilon
_{2}=1$ here. For the relatively weak hopping between the two QDs
with respect to the couplings of QD-1 with two metallic electrodes,
i.e., $J<\Gamma $, Figures 2 and 3 show the first four cumulants of
zero-frequency current fluctuation as a function of bias voltage for
the cases of considering and without considering quantum coherence
at different values of\quad $\Gamma $, respectively. Considering
only the diagonal elements of the reduced density matrix, the
super-Poissonian noise cannot be observed; but the corresponding
Fano factors, which do not depend on the $\Gamma $ value, have the
same values, see the Fig. 3. This characteristic can be understood
in terms of
the specific form of $\mathcal{L}_{\chi }$. For the considered case of $%
\Gamma _{L}=\Gamma _{R}=\Gamma $, the matrix $\mathcal{L}_{\chi }$
can be rewritten as $\Gamma \mathcal{L}_{\chi }^{\prime }$, and the
corresponding $k
$-order cumulant is given by $C_{k}/\Gamma $, so that the Fano factor of $k$%
-order cumulant, which are defined as $C_{k}/C_{1}$, does not depend on the $%
\Gamma $ value, see the Fig. 3. As a result, the off-diagonal
elements of the reduced density matrix should be considered in the
numerical calculation. In particular, the quantum coherence, which
originates from the two kinds of coherent singly-occupied
eigenstates $\left\vert \Psi _{1,\sigma }^{+}\right\rangle $ and
$\left\vert \Psi _{1,\sigma }^{-}\right\rangle $ determined by the
hopping parameter $J$, plays a crucial role in determining whether
the super-Poissonian noise occurs for the $\Gamma >J$ case, where
this ratio of $\Gamma $ to $J$ should be greater than a certain
value (about $3$), see the Figs 2(f), 2(g) and 2(h).\ This
characteristic of super-Poissonian noise can be understood in terms
of the interplay of quantum coherence and effective competition
between correlated electron transport channels, which is a new
occurrence-mechanism of
super-Poissonian noise and not revealed in the previous studies\cite%
{WangSK,Djurica,Xue01,Xue02,Xue03,Xue04,Aghassi}. For this
super-Poissonian noise bias voltage range, electrons can tunnel from
the left metallic electrode into and then tunnel out of the DQD
system to the right electrode via the transition between the
singly-occupied and empty eigenstates. In the case of $\Gamma \gg
J$, when the conduction-electron from the left electrode tunnels
from the QD-1 to the QD-2, this electron in the QD-2 will remain for
a relatively long time, then again tunnels into the QD-1 and out of
the double-QD system to the right metallic electrode. These indirect
electron tunneling processes will lead to electron tunneling is
blocked since this DQD system can occupy only one electron and form
the slow electron transport channels, whereas the direct tunneling
processes through QD-1 form the corresponding fast transport
channels. The interplay of quantum coherence and effective
competition between the fast and slow correlated electron transport
channels eventually results in the formation of super-Poissonian
noise, and the super-Poissonian behavior will be more obvious for
the relatively large ratio of $\Gamma $ to $J$, see the Figs. 2(f),
2(g) and 2(h).

However, the super-Poissonian noise value will be decreased to a
sub-Poissonian value as expected for the relatively strong hopping, i.e., $%
J>\Gamma $. This originates from the fact that with increasing the ratio of $%
J$ to $\Gamma $ especially for the $J\gg \Gamma $ case the
conduction-electron can tunnel back and forth between the two QDs
that leads to the indirect electron tunneling processes are almost
equivalent to direct electron tunneling processes, so that the
effective fast and slow correlated electron transport channels
cannot be formed, and the quantum coherence of the two kinds of
coherent singly-occupied eigenstates will be weakened, namely, the
diagonal elements of the reduced density matrix play a major role in
the electron tunneling processes, see the Fig. 4. Consequently, the
corresponding Fano factors decrease to the sub-Poissonian values.
Another important finding is that the high-order cumulants, e.g.,
the skewness and kurtosis, are more sensitive to the quantum
coherence of this considered QDs
system, which is independent of the relative magnitudes of $J$ and $\Gamma $%
,\ see the Figs. 2(g) and 2(h), and Fig. 4. This characteristic can
be used to detect the quantum coherence of quantum systems, which is
contrary to the conclusion of the Ref. \cite{Pala}.

Finally, we study the effect of the ground state (which can be tuned
by a gate voltage $V_{g}$) of the present QDs system on the
super-Poissonian behaviors, here the corresponding parameters of $J$
and $\Gamma $ are chosen as $\Gamma =0.005$ and $J=0.001$. The
empty, singly-occupied and
doubly-occupied eigenstates are given by%
\begin{equation}
\left\{
\begin{array}{c}
\epsilon _{0,0}=0 \\
\epsilon _{1,\sigma }^{\pm }=\epsilon \pm J \\
\epsilon _{\sigma ,\sigma ^{\prime }}=2\epsilon +U_{12}%
\end{array}%
\right. ,  \label{eigenstates}
\end{equation}%
where $\epsilon _{1}=\epsilon _{2}=\epsilon $. Therefore, for the
$\epsilon
>0$ case the ground state is the electronic unoccupied eigenstate; but for $%
\epsilon <0$ the ground states are the singly-occupied eigenstates $%
\left\vert \Psi _{1,\sigma }^{\pm }\right\rangle $ or
doubly-occupied eigenstates $\left\vert \Psi _{\sigma ,\sigma
^{\prime }}\right\rangle $ since $\epsilon \gg J$. In the two cases
of $\epsilon >0$, and $\epsilon <0$ and $2\epsilon +U_{12}>0$, the
first current-step induced by the transition processes between the
singly-occupied and empty eigenstates, the super-Poissonian noise in
this bias voltage, which originates from the same mechanism as
mentioned above, can occur and the threshold value of the
super-Poissonian noise bias voltage depends on $\epsilon $. This
characteristic can be used to tune the super-Poissonian noise bias
voltage range by adjusting the two independent gate electrodes for
each dot, and suggests a tunable super-Poissonian noise device, see
the Fig. 5. As for the $\epsilon <0$ and $2\epsilon +U_{12}>0$ case,
the first current-step induced by the transition processes between
the doubly-occupied and singly-occupied eigenstates. For the
considered $\Gamma /J=5$ case, when one of the two conduction
electrons tunnels out the DQD system to the right electrode, the
singly-occupied eigenstates will relax to the state of $\left\vert
\sigma \right\rangle _{1}\left\vert 0\right\rangle _{2}$ or
$\left\vert 0\right\rangle _{1}\left\vert \sigma \right\rangle _{2}$
because the system needs to be formed a new doubly-occupied
eigenstate. If the other conduction electron in the QD-1, i.e.,
$\left\vert \sigma \right\rangle _{1}\left\vert 0\right\rangle
_{2}$, that will remain for a relatively long time until this
electron tunnels from the QD-1 to the QD-2, then electron from the
left electrode can tunnel into the QD-1. This process will result in
electron tunneling is blocked and form the slow transport channels;
whereas for the case of the other conduction electron in the QD-2,
electron from the left electrode can direct tunnel into the QD-1 and
form a new doubly-occupied eigenstate, so that electron tunneling
processes can continues until the singly-occupied eigenstates relax
again to the $\left\vert \sigma \right\rangle _{1}\left\vert
0\right\rangle _{2}$ state, which will form the corresponding fast
correlated transport channels. The interplay of quantum coherence
and effective competition between these fast and slow correlated
transport channels is responsible for the super-Poissonian noise
behavior, see the short-dashed lines in the Figs. 5(b), 5(c) and
5(d). Furthermore, the skewness and kurtosis for the $\epsilon <0$
and $2\epsilon +U_{12}>0$ case, respectively, show a very large
negative value relative to the two cases of $\epsilon >0$, and
$\epsilon <0$ and $2\epsilon +U_{12}>0$, see the Figs. 5(c) and
5(d). This characteristic can be used to identify whether the ground
state is doubly-occupied eigenstate.

\section{Conclusions}

We have studied theoretically the full counting statistics of
electron transport through the side-coupled DQD system above the
sequential tunneling threshold. The high-order cumulants of
transport current are found to be more sensitive to the quantum
coherence than the average current. This characteristic can be used
to probe the quantum coherence of the considered side-coupled DQD
system. Especially, the quantum coherence of this DQD system plays a
crucial role in determining whether the super-Poissonian noise
occurs in the weak inter-dot hopping coupling regime, e.g.,
super-Poissonian noise is observed when the ratio of the inter-dot
hopping coupling to dot-lead coupling is smaller than a certain
value, and the corresponding values of super-Poissonian noise can be
relatively enhanced when considering conduction electron spin.
Moreover, this super-Poissonian noise bias range depends on the
singly-occupied eigenstates of the system, which thus suggests a
tunable super-Poissonian noise device. The super-Poissonian noise
characteristics can be qualitatively attributed to the interplay of
quantum coherence and effective competition between fast and slow
transport channels.

\section*{Acknowledgments}

This work was supported by the National Nature Science Foundation of
China (Grant Nos. 11204203 and 61274089), and the Youth Foundation
of Taiyuan University of Technology (Grant No. 2012L039).

\newpage

Figure Caption:

Fig. 1 Schematic representation of the considered side-coupled DQD
system coupled to two metallic electrodes.

Fig. 2 (Color online) The first four cumulants of zero-frequency
current fluctuation versus bias voltage for different\quad coupling
of the side-coupled DQD system with two metallic electrodes. Here,
the off-diagonal elements of the reduced density matrix are
considered. (a), (b), (c) (d) for the spinless Hamiltonian; and (e),
(f), (g), (h) for the spin Hamiltonian. The system parameters:
$\epsilon _{1}=\epsilon _{2}=1$, $U_{12}=5$, $J=0.001$ and
$k_{B}T=0.1$.

Fig. 3 (Color online) The first four cumulants of zero-frequency
current fluctuation versus bias voltage for different\quad coupling
of the side-coupled DQD system with two metallic electrodes. Here,
the diagonal elements of the reduced density matrix are only
considered. (a), (b), (c) (d) for the spinless Hamiltonian; and (e),
(f), (g), (h) for the spin Hamiltonian.. The other system parameters
are the same as in Fig. 2.

Fig. 4 (Color online) The first four cumulants of zero-frequency
current fluctuation versus bias voltage for the two cases for
considering off-diagonal elements and only considering diagonal
elements at fixed coupling of the side-coupled DQD system with two
metallic electrodes, respectively. (a), (b), (c) (d) for $J=0.005$
and $\Gamma _{\uparrow}=\Gamma _{\downarrow }=0.001$; and (e), (f), (g), (h) for $J=0.01$ and $%
\Gamma _{\uparrow }=\Gamma _{\downarrow }=0.001$. The other system
parameters are the same as in Fig. 2.

Fig. 5 (Color online) The first four cumulants of zero-frequency
current
fluctuation versus bias voltage for different values of $\epsilon $ $%
(\epsilon _{1}=\epsilon _{2}=\epsilon )$, where $J=0.001$ and
$\Gamma _{L\uparrow }=\Gamma _{L\downarrow }=\Gamma _{R\uparrow
}=\Gamma _{R\downarrow }=0.005$. The other parameters are the same
as in Fig. 2.

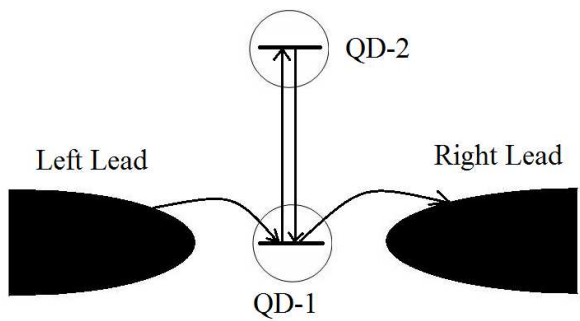
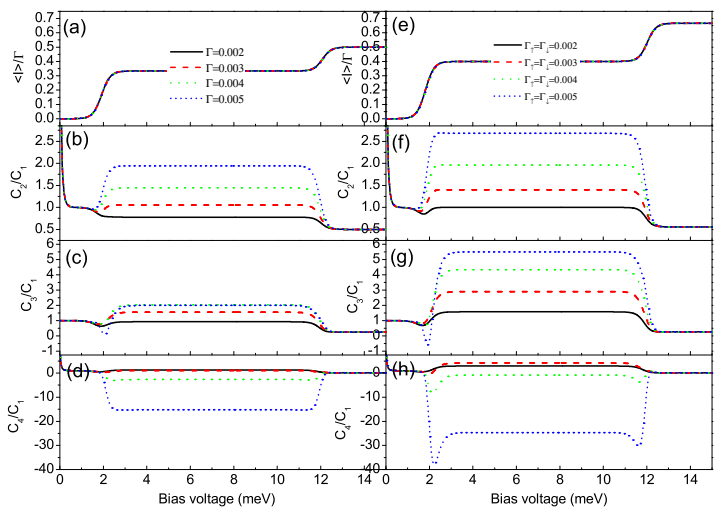
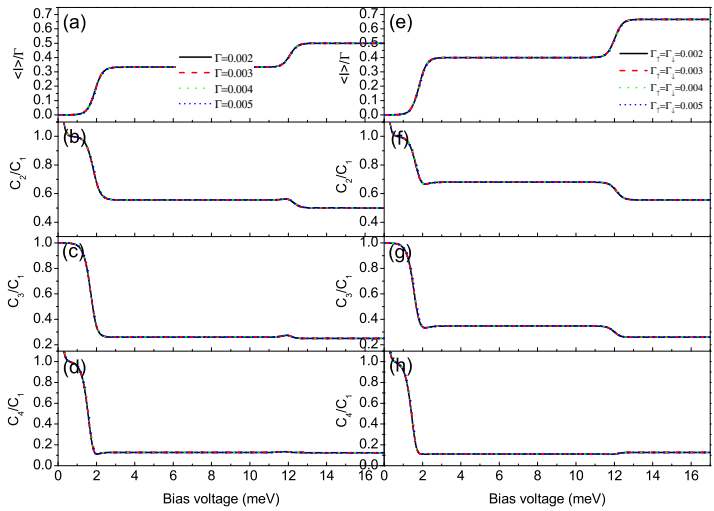
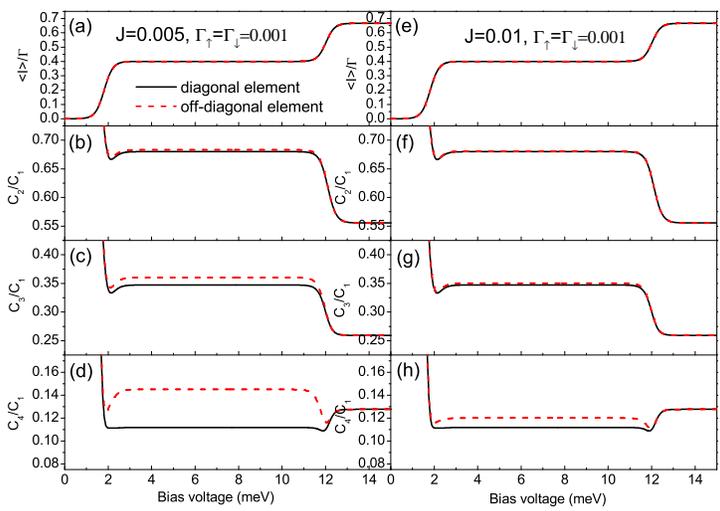
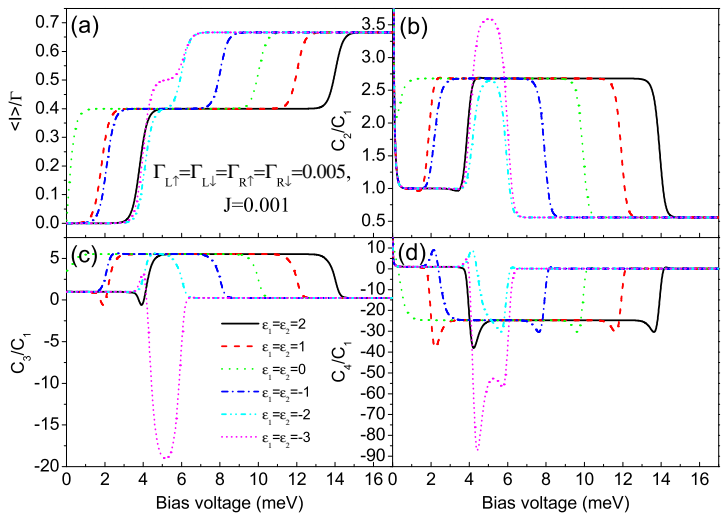
\end{document}